\documentclass[12pt]{amsart}
\usepackage{amsfonts,amssymb}
\usepackage{amsrefs}
\makeatletter
\def\thickhrulefill{\leavevmode \leaders \hrule height 2pt\hfill \kern \z@}
\makeatother

\def\mytitle #1{\raisebox{0pt}{
\vbox{\hbox to 0.99\textwidth{ \strut \sc #1
\hfil \shadowbox{\thepage} 
}\hrule height 2pt}}}

\newcommand{\eps}{\varepsilon}

\newcommand{\G}{\mathcal G}
\newcommand{\Z}{\mathbb{Z}}

\renewcommand{\S}{\mathcal S}
\newcommand{\F}{\mathcal F}

\renewcommand{\phi}{\varphi}

\renewcommand{\subset}{\subseteq}

\def\now{
\ifnum\time<60
          12:\ifnum\time<10 0\fi\number\time am
          \else
            \ifnum\time>719\chardef\a=`p\else\chardef\a=`a\fi
          \hour=\time
          \minute=\time
          \divide\hour by 60 
          \ifnum\hour>12\advance\hour by -12\advance\minute by-720 \fi
          \number\hour:%
          \multiply\hour by 60 
          \advance\minute by -\hour
          \ifnum\minute<10 0\fi\number\minute\a m\fi}
\newcount\hour
\newcount\minute
\numberwithin{equation}{section}         
\newtheorem{thm}{Theorem}[section]

\newtheorem{defn}[thm]{Definition}
\theoremstyle{remark}
\newtheorem{remark}{Remark}[section]

\makeatletter
\renewcommand{\@evenhead} 
{\raisebox{0pt}[\headheight][0pt]{
\vbox{\hbox to 0.99\textwidth{ {\small \thepage}\hfil \strut
{\sc\small A.C.D.~van Enter, E.A.~Verbitskiy}
\hfil 
}\hrule}} }

\renewcommand{\@oddhead} 
{\raisebox{0pt}[\headheight][0pt]{
\vbox{\hbox to 0.99\textwidth{ \hfil \strut {\sc\small  Variational Principle  for Generalized Gibbs
Measures }\hfil{\small \thepage}}\hrule}} }

\renewcommand{\@oddfoot}{}
\renewcommand{\@evenfoot}{}

\newcommand{\l@abcd}[2]{\hbox to\textwidth{#1\dotfil #1}}

\makeatother

\begin{document}
\title{On the  Variational Principle  for Generalized Gibbs
Measures}
\author[A.C.D.van Enter]{{\sc Aernout van Enter}}
\address{A.C.D.van Enter, Institute for Theoretical Physics, University of Groningen, Nijenborgh 4, 9747 AG Groningen,
The Netherlands}
\email{a.c.d.van.enter@phys.rug.nl}
\author[E.A. Verbitskiy]{{\sc Evgeny Verbitskiy}}
\address{E.A. Verbitskiy, Philips Research Laboratories, Prof. Holstlaan 4 (WO 2), 5656 AA Eindhoven, The Netherlands}
\email{evgeny.verbitskiy@philips.com}

\begin{abstract}
We present  a novel  approach to establishing  the variational  principle for
Gibbs  and generalized  (weak  and  almost) Gibbs  states.  Limitations of  a
thermodynamic formalism  for generalized Gibbs  states will be  discussed.  A
new  class  of  intuitively  Gibbs  measures is  introduced,  and  a 
typical  example  is studied.   Finally,  we present  a  new  example of  a
non-Gibbsian measure arising from an industrial application.

\end{abstract}

\maketitle
\section{Introduction}

Gibbs  measures,   defined  as  solutions   of  the  Dobrushin-Lanford-Ruel\-le
equations,  can  equivalently  be  defined  as  solutions  of  a  variational
principle (at least when they are translation invariant).

Such a  variational principle states  that when we  take as a base  measure a
Gibbs measure for some potential,  or more generally, for some specification,
other Gibbs measures for the same potential (specification) are characterized
by having a zero relative entropy density with respect to this base measure.

If the base measure is not a Gibbs measure, such a statement need not be true
anymore.  The  construction  of  Xu   \cite{xu}  provides  an  example  of  a
``universal''  ergodic  base measure,  such  that  all translation  invariant
measures  have zero  entropy  density with  respect  to it.  Even within  the
reasonably well-behaved class of ``almost Gibbs'' measures there are examples
such that   certain Dirac measures  have 
zero entropy density  with respect to it,
\cite{ES}; for a similar weakly Gibbsian example see \cite{maesetall2}.   
One  can, however,  to  some  extent circumvent  this
problem   by   requiring  that   both   measures   share  sufficiently   many
configurations in their support.

For  almost  Gibbs  measures,   the  measure-one  set  of  good  (continuity)
configurations have the property that  they can shield off any influence from
infinity. On the other hand for  the strictly larger class of weakly Gibbsian
measures, it may suffice that  most, but not necessarily all, influences from
infinity are blocked by the ``good'' configurations.

The situation with respect to  the variational principle between the class of
almost Gibbs measures  is much better than with respect to  the class of weak
Gibbs measures \cite{KLR}.

On the one  hand, one expects that a variational  principle might hold beyond
the   class  of   almost   Gibbs  measures.    For  example,   infinite-range
unbounded-spin systems lack  the almost Gibbs property (due  to the fact that
for a configuration of  sufficiently increasing spins the interaction between
the origin  and infinity is  never negligible, whatever happens  in between),
but a  variational principle  for such  models has been  found; on  the other
hand, the analysis of K\"ulske for  the random field Ising model implies that
one really needs  some extra conditions, or the  variational principle can be
violated.

The paper  is organised  as follows. After  recalling some basic  notions and
definitions, we  discuss Goldstein's construction  of a specification  for an
arbitrary  translation  invariant measure.  In  Section  3,  we consider  two
measures  $\nu$  and  $\mu$  such  that $h(\nu|\mu)=0$  and  we  formulate  a
sufficient condition  for $\nu$ to  be consistent with a  given specification
$\gamma$ for $\mu$. We also consider a general situation of $h(\nu|\mu)=0$ 
and recover 
a result of F\"ollmer. The new  sufficient condition is clarified in the case
of  almost Gibbs  measures in  Section  4, and  for a  particular weak  Gibbs
measure in Section 5. We also introduce a new class of intuitively weak Gibbs
measures.   In Section 6,  we present  an example  of a  non-Gibbsian measure
arising in industrial setting: magnetic and optical data storage.

{\bf Acknowledgment.} We are grateful to Frank den Hollander, Christ\-of K\"ulske,
Jeff Steif, and Frank Redig for useful discussions.

\section{Specifications and Gibbs measures}
\subsection{Notation}
We work with spin systems  on the lattice $\mathbb Z^d$, i.e., configurations
are elements of the product space $\mathcal A^{\mathbb Z^d}$, where $\mathcal
A$  is  a   finite  set  (alphabet).   The  configuration   space  $\Omega  =
\mathcal{A}^{\Z^d}$ is  endowed with the  product topology, making it  into a
compact metric  space.  Configurations will denoted
by lower-case Greek letters. The set  of finite subsets  
of $\Z^d$ is  denoted by $\mathcal S$.

 For $\Lambda\in\mathcal S$ we put $\Omega_\Lambda=\mathcal{A}^\Lambda$.  For
 $\sigma\in\Omega$,           and           $\Lambda\in\mathcal           S$,
 $\sigma_\Lambda\in\Omega_\Lambda$  denotes the restriction  of $\sigma  $ to
 $\Lambda$.   For  $\sigma$,  $\eta$  in  $\Omega$,  $\Lambda\in\mathcal  S$,
 $\sigma_{\Lambda}\eta_{\Lambda^c}$ denotes the configuration coinciding with
 $\sigma$  on  $\Lambda$, and  $\eta$  on  $\Lambda^c$.  For  $\Lambda\subset
 \mathbb Z^d$,  $\mathcal F_\Lambda$ denote the  $\sigma$-algbra generated by
 $\{\sigma_x|\, x\in \Lambda\}$.

For two translation invariant probability measures $\mu$ and $\nu$, define
$$ H_{\Lambda} (\nu|\mu) = H(\nu_\Lambda|\mu_\Lambda) = \sum_{\sigma_\Lambda}
\nu(\sigma_\Lambda) \log \frac { \nu(\sigma_\Lambda)} { \mu(\sigma_\Lambda)},
$$ if  $\nu_\Lambda$ is absolutely continuous with  respect to $\mu_\Lambda$,
and $H_{\Lambda} (\nu|\mu)=+\infty$, otherwise.

The  relative entropy  density $h(\nu|\mu)$  is defined  (provided  the limit
exists) as
$$ h(\nu|\mu)=\lim_{n\to\infty} \frac 1{|\Lambda_n|} H_{\Lambda_n}(\nu|\mu),
$$ where $\{\Lambda_n\}$ is a  sequence of finite subsets $\mathbb Z^d$, with
$\Lambda_n\nearrow  \mathbb Z^d$  as $n\to\infty$  in van  Hove  sense.  For
example, one can take $\Lambda_n =[-n,n]^d$.

A potential  $U=\{ U(\Lambda,\cdot)\}_{\Lambda\in\mathcal S}$ is  a family of
functions indexed by  finite subsets of $\mathbb Z^d$  with the property that
$U(A,\omega)$ depends  only on $\omega_\Lambda$.  A Hamiltonian $H_\Lambda^U$
is defined by
$$     H_\Lambda^U(\sigma)     =     \sum_{\Lambda'\cap\Lambda\ne\varnothing}
 U(\Lambda',\sigma).
$$ The Hamiltonian $H_\Lambda^U$ is said  to be convergent in $\sigma$ if the
sum  on  the  right  hand   side  is  convergent.   The  Gibbs  specification
$\gamma^U=\{\gamma_\Lambda^U\}_{\Lambda\in\mathcal S}$ is defined by
$$     
\gamma_\Lambda^U(    \omega_\Lambda|\sigma_{\Lambda^c})     =    
\frac
{\exp\Bigl(-H_\Lambda^U(\omega_\Lambda\sigma_{\Lambda^c})              \Bigr)}
{
\sum\limits_{\tilde\omega_\Lambda\in\Omega_\Lambda}
\exp\Bigl(-H_\Lambda^U(\tilde\omega_\Lambda\sigma_{\Lambda^c}) \Bigr)}
$$    provided     $H_\Lambda^U$    is    convergent     in    every    point
$\tilde\omega_\Lambda\sigma_{\Lambda^c}$,
$\tilde\omega_\Lambda\in\Omega_\Lambda$.   Formally, if $H_\Lambda^U$  is not
convergent   in  every   point   $\omega\in\Omega$,  $\gamma^U$   is  not   a
specification  in  the  sense  of standard  Definition  \ref{def_specif_org}.
Nevertheless, in  many cases (e.g., weakly Gibbsian  measures, see Definition
\ref{defn_weakgibbs} below)  $\gamma^U$ can still  be viewed as a  version of
conditional probabilities for $\mu$:
$$       \mu(\omega_\Lambda|\sigma_{\Lambda^c})      =      \gamma^U_\Lambda(
\omega_\Lambda|\sigma_{\Lambda^c})\quad \text{($\mu$-a.s.)}.
$$

\subsection{Specifications}
\begin{defn}\label{def_specif_org} A family of probability kernels $\gamma = \{
\gamma_\Lambda\}_{\Lambda\in\mathcal S}$ is called a specification
if
\begin{itemize}
\item[a)] $\gamma_\Lambda(F|\cdot)$ is $\mathcal
F_{\Lambda^c}$-measurable for all $\Lambda\in\mathcal S$ and $F\in
\mathcal F$; \item[b)] $\gamma_{\Lambda}(F|\omega) = \mathbb
I_F(\omega)$ for all $\Lambda\in\mathcal S$ and
 $F\in\mathcal F_{\Lambda^c}$;
\item[c)] $\gamma_{\Lambda'} \gamma_{\Lambda} =\gamma_{\Lambda'}$
whenever $\Lambda\subset \Lambda'$, and where
$$
\bigl(\gamma_{\Lambda'} \gamma_{\Lambda}\bigr)(F|\omega) = \int
\gamma_{\Lambda}( F |\eta) \gamma_{\Lambda'}( d\eta |\omega).
$$
\end{itemize}
\end{defn}

\begin{defn}\label{def_Gibbs_spec} A probability measure $\mu$ is
called consistent  with a specification $\gamma$  (denoted by $\mu\in\mathcal
G(\gamma)$) if for every bounded measurable function $f$ one has
\begin{equation}\label{DLR_spec}
\int f\,d\mu =\int \gamma_\Lambda (f)\, d\mu.
\end{equation}
\end{defn}
If $\mu$ is consistent with  the specification $\gamma$, then $\gamma$ can be
viewed   as  a  version   of the 
conditional   probabilities  of   $\mu$,  since
(\ref{DLR_spec}) implies that for any finite $\Lambda$
$$   \gamma_\Lambda(   A|\omega)   =\mathbb   E_\mu(   \mathbb   I_A|\mathcal
F_{\Lambda^c})(\omega)\quad (\text{$\mu$-a.s.}),
$$ where $\mathcal F_{\Lambda^c}$  is the $\sigma$-algebra generated by spins
outside $\Lambda$.

Definition   \ref{def_specif_org}   requires   that   for   every   $\omega$,
$\gamma_\Lambda(\cdot|\omega)$ is a probability measure on $\F$, and that the
consistency condition  (c) is satisfied  for {\bf all}  $\omega\in\Omega$. In
fact, when dealing with the weakly Gibbs measures, these requirements are too
strong.     Definition     \ref{def_specif_org}     can    be     generalized
\cite[p. 16]{preston}, and this form is probably more suitable for the weakly
Gibbsian formalism.

\subsection{Construction of specifications} In \cite{goldstein}, Goldstein showed that every measure
has a specification.  In other words, for every measure  $\mu$ there exists a
specification $\gamma$ in the  sense of definition \ref{def_specif_org}, such
that   $\mu\in\mathcal  G(\gamma)$.   Let  us   briefly   recall  Goldstein's
construction.

Suppose  $\mu$  is  a  probability  measure  on  $\Omega=A^{\Z^d}$,  and  let
$\{\Lambda_n\}$,  $\Lambda_n\in\mathcal S$,  be an  increasing  sequence such
that  $\cup_n  \Lambda_n=\Z^d$.  For  a  finite  set  $\Lambda \in\Z^d$,  and
arbitrary          $\eta_\Lambda\in          A^{\Lambda}$,         $\omega\in
A^{\Z^d\setminus\Lambda}$, define
$$  \mu(\eta_\Lambda| \omega_{\Lambda^c} )  := \mu(  [\eta_\Lambda]| \mathcal
F_{\Lambda^c})(\omega),
$$   
where   $[\eta_\Lambda]=   \{\zeta\in\Omega:   \,\,   \zeta|_\Lambda   =
\eta_\Lambda\}$. By the martingale convergence theorem
\begin{equation}\label{conv1}
 \mu(  [\eta_\Lambda]|  \mathcal  F_{\Lambda^c})(\omega) =  \lim_{n\to\infty}
 \mu(   \eta_\Lambda|  \omega_{\Lambda_n\setminus\Lambda})   \quad  \text{for
 }\mu-a.e. \ \omega.
\end{equation}
The sequence on the right hand side of (\ref{conv1}) is defined by elementary
conditional probabilities:
$$      \mu(      \eta_\Lambda|     \omega_{\Lambda_n\setminus\Lambda})=\frac
{\mu(\eta_\Lambda       \omega_{\Lambda_n\setminus\Lambda})}      {      \mu(
\omega_{\Lambda_n\setminus\Lambda})}
$$ 
Define
\begin{equation}\label{GLambda}
G_{\Lambda} =\{ \omega: \, \text{the limit on RHS of (\ref{conv1})
exists for all }\, \eta_\Lambda\in\Omega_\Lambda\},
\end{equation}
and denote this limit by $p_\Lambda(\eta|\omega)$. For
$\Lambda\subset \Lambda_n$, let
$$
Q_{\Lambda}^{\Lambda_n} = \{\omega\in G_{\Lambda_n}: \,
\sum_{\eta_{\Lambda}\in\Omega^{\Lambda}}
p_{\Lambda_n}(\eta_{\Lambda}\omega_{\Lambda_n\setminus\Lambda}| \omega)>0
\}.
$$
Finally, let
$$
\mathcal H_{\Lambda} = \bigcup_{n}\bigcap_{j=n}^{\infty}
Q_{\Lambda}^{\Lambda_j},
$$
and define $\gamma_{\Lambda}$ by
\begin{equation}\label{spec_gold}
\gamma_{\Lambda}(\eta|\omega) = \begin{cases} 
p_{\Lambda}(\eta|\omega), &\text{if }\omega\in \mathcal H_{\Lambda}\\
(|\Lambda||\mathcal A|)^{-1}, &\text{if }\omega \in \mathcal H_{\Lambda}^c
\end{cases}.
\end{equation}

\begin{thm} The family $\gamma=\{\gamma_{\Lambda}\}_{\Lambda\in\mathcal S}$ given by (\ref{spec_gold})
is a specification, and $\mu\in\mathcal G(\gamma)$.
\end{thm}

 Suppose $\gamma$ is a specification,
and $\mu$ is cosnsistent with $\gamma$. Therefore
$$
\mu(\cdot| \omega_{\Lambda^c} ) =\gamma_\Lambda(\cdot| \omega_{\Lambda^c} )\quad\mu-a.s.
$$
Taking (\ref{conv1}) into account we conclude that
\begin{equation}\label{conv3}
\gamma_\Lambda(\eta_\Lambda| \omega_{\Lambda^c} )=\lim_{n\to\infty} \mu( \eta_\Lambda| \omega_{\Lambda_n\setminus\Lambda})
\end{equation}
for all  $\eta_\Lambda$ and $\mu$-almost all $\omega$.   An important problem
for establishing the variational principle for generalized Gibbs measures, is
determining the set of  configurations where the convergence in (\ref{conv3})
takes place.

\section{Properly supported measures}
\begin{thm}\label{main-res}
Let $\mu$  be a measure  consistent with the specification  $\gamma$. Suppose
that $\nu$ is  another measure such that $h(\nu|\mu)=0$.  If for $\nu$-almost
all $\omega$
\begin{equation}\label{newcond}
\mu(\xi_\Lambda|            \omega_{\Lambda_n\setminus\Lambda})           \to
\gamma(\xi_\Lambda|\omega_{\Lambda^c}),
\end{equation}
then   $\nu$   is  consistent   with   the   specification  $\gamma$,   i.e.,
$\nu\in\G(\gamma)$.
\end{thm}

\begin{remark}
Note   that   by   the   dominated  convergence   theorem,   convergence   in
(\ref{newcond}) is also in $L_1(\nu)$.
\end{remark}

\begin{remark} If $\mu$ is an almost Gibbs measure for the specification $\gamma$,
and $\nu$ is a measure concentrating on the points of continuity of $\gamma$, i.e.,
$\nu(\Omega_\gamma)=1$, then (\ref{newcond}) holds, see the proof below.
\end{remark}

\begin{proof}[Proof of Theorem \ref{main-res}]
Suppose  that  $h(\nu|\mu)=0$.  Then  \cite[Theorem 15.37]{georgii}  for  any
$\eps>0$  and  any  finite  set  $\Lambda$,  and every  cube  $C$  such  that
$\Lambda\subset C$, there exists $\Delta$, $C\subset \Delta$ such that
\begin{equation}\label{densestim}
\mu( |f_\Delta -f_{\Delta\setminus\Lambda}|) <\eps,
\end{equation}
where for any finite set $V$, $f_V$ is the density  of $\nu|_V$ with respect to 
$\mu|_V$:
$$
f_V (\omega_V) = \frac {\nu(\omega_V)}{\mu(\omega_V)}.
$$
Rewrite (\ref{densestim}) as follows:
\begin{align*}
\mu( |f_\Delta -f_{\Delta\setminus\Lambda}|)&=
\sum_{\eta_\Lambda,\omega_{\Delta\setminus\Lambda}}
\mu(\eta_\Lambda\omega_{\Delta\setminus\Lambda}) \Bigl|
\frac {\nu(\eta_\Lambda\omega_{\Delta\setminus\Lambda})}{\mu(\eta_\Lambda\omega_{\Delta\setminus\Lambda})}
-
\frac {\nu( \omega_{\Delta\setminus\Lambda})}{\mu( \omega_{\Delta\setminus\Lambda})}
\Bigr|\\
&=\sum_{\omega_{\Delta\setminus\Lambda}} \nu( \omega_{\Delta\setminus\Lambda})
\Bigl\{\sum_{\eta_{\Lambda}}\Bigl|
\frac {\nu(\eta_\Lambda\omega_{\Delta\setminus\Lambda})}{\nu( \omega_{\Delta\setminus\Lambda})}
-
\frac {\mu( \eta_\Lambda\omega_{\Delta\setminus\Lambda})}{\mu( \omega_{\Delta\setminus\Lambda})}
\Bigr|\Bigr\}\\
&=\sum_{\omega_{\Delta\setminus\Lambda}} \nu( \omega_{\Delta\setminus\Lambda})
||  \nu_{\Lambda}( \cdot |\omega_{\Delta\setminus\Lambda})  -
\mu_{\Lambda}( \cdot |\omega_{\Delta\setminus\Lambda})||_{TV}\\
&= \mathbb E_{\nu} ||  \nu_{\Lambda}( \cdot |\omega_{\Delta\setminus\Lambda})  -
\mu_{\Lambda}( \cdot |\omega_{\Delta\setminus\Lambda})||_{TV},
\end{align*}
where $||\cdot||_{TV}$ is the total variation norm.

A measure $\nu$ is consistent with the specification $\gamma$ if
$$
 \nu_{\Lambda}( \cdot |\omega_{\Lambda^c})=\gamma_{\Lambda}( \cdot |\omega_{\Lambda^c}), \quad \nu-a.e.,
$$
or, equivalently,
$$
\mathbb E_{\nu} ||  \nu_{\Lambda}( \cdot |\omega_{\Lambda^c})  -
\gamma_{\Lambda}( \cdot |\omega_{\Lambda^c})||_{TV} =0.
$$
Obviously one has
\begin{align*}
\mathbb E_{\nu} ||  \nu_{\Lambda}( \cdot |\omega_{\Lambda^c})  -
\gamma_{\Lambda} &( \cdot |\omega_{\Lambda^c})||_{TV}\\
\le&\,
\mathbb E_{\nu} ||  \nu_{\Lambda}( \cdot |\omega_{\Lambda^c})  -
\nu_{\Lambda}( \cdot |\omega_{\Delta\setminus\Lambda})||_{TV} \\
&\quad+\mathbb E_{\nu} ||  \nu_{\Lambda}( \cdot |\omega_{\Delta\setminus\Lambda})  -
 \mu_{\Lambda}( \cdot |\omega_{\Delta\setminus\Lambda}) ||_{TV} \\
&\quad\quad+\mathbb E_{\nu} ||  \mu_{\Lambda}( \cdot |\omega_{\Delta\setminus\Lambda})  -
\gamma_{\Lambda}( \cdot |\omega_{\Lambda^c})||_{TV}.
\end{align*}
By the martingale convergence theorem, the first term
$$
\mathbb E_{\nu} ||  \nu_{\Lambda}( \cdot |\omega_{\Delta\setminus\Lambda})  -
\nu_{\Lambda}( \cdot |\omega_{\Lambda^c})||_{TV} \to 0,\quad\text{as }\Delta\nearrow\Z^d.
$$
The second  term
tends to 0 due to (\ref{densestim}), and the third term tends to zero because
of our assumptions.\end{proof}

\subsection{Weakly Gibbs measures which violate the Variational Principle}
Disordered systems studied extensively by K\"ulske \cite{Kuelske1, Kuelske2} 
provide a counterexample to the variational principle for weakly 
Gibbs measures. In \cite{KLR}, K\"ulske, Le Ny and Redig showed
that there exist two weakly Gibbs measures $\mu^+$, $\mu^-$ such that
\begin{equation}\label{ent_dens_sym}
h(\mu^+|\mu^-) = h(\mu^-|\mu^+) = 0,
\end{equation}
but $\mu^+$ is not consistent with a  (weakly) 
Gibbsian specification $\gamma^-$ for $\mu^-$, and vice versa.
The novelty and beauty of their Random Field Ising Model 
example lies in the fact  that {\em both} measures are non-trivial and
the relation in  (\ref{ent_dens_sym})  is symmetric.
As already mentioned above, previous almost Gibbs (\cite{ES})
and weakly Gibbs \cite{maesetall2}   example 
violating the variational principle
satisfied $h(\delta| \mu)=0$ with $\delta$ being a Dirac measure.

What  happens in the situation when $h(\nu|\mu)=0$?
Suppose $\nu$ is consistent with the specification $\tilde\gamma$. Then
\begin{align*}
\mathbb E_{\nu} ||  \mu_{\Lambda}( \cdot |\omega_{\Delta\setminus\Lambda})  -
\tilde\gamma_{\Lambda}( \cdot |\omega_{\Lambda^c})||_{TV}\le&
\mathbb E_{\nu} ||  \mu_{\Lambda}( \cdot |\omega_{\Delta\setminus\Lambda})  -
 \nu_{\Lambda}( \cdot |\omega_{\Delta\setminus\Lambda}) ||_{TV} \\
&+\mathbb E_{\nu} ||  \nu_{\Lambda}( \cdot |\omega_{\Delta\setminus\Lambda})  -
\tilde\gamma_{\Lambda}( \cdot |\omega_{\Lambda^c})||_{TV}.
\end{align*}
Again, the first term on the right hand side tends to zero because $h(\nu|\mu)$ =0, and
the second term tends to zero, because of the martingale convergence theorem.
Therefore
$$
||  \mu_{\Lambda}( \cdot |\omega_{\Delta\setminus\Lambda})  -
\tilde\gamma_{\Lambda}( \cdot |\omega_{\Lambda^c})||_{TV} \to 0,
$$
in $L_1(\nu)$, and since  the total variation between any two measures is 
always bounded by $2$,
we have that
$$
 \mu_{\Lambda}( \cdot |\omega_{\Delta\setminus\Lambda})  \to
\tilde\gamma_{\Lambda}( \cdot |\omega_{\Lambda^c}), \quad \nu-a.s.
$$
Since $\tilde\gamma$ is a specification for $\nu$, and since the 
"infinite" conditional probability for $\mu$ are
defined as the limits of finite conditional probabilities (provided 
the limits exist) we conclude that
\begin{equation}\label{follmer_reform}
 \mu_{\Lambda}( \cdot |\omega_{\Lambda^c})=
\nu_\Lambda( \cdot |\omega_{\Lambda^c}), \quad \nu-a.s.
\end{equation}
In fact  (\ref{follmer_reform}) was obtained in a different way 
earlier by F\"ollmer in \cite[Theorem 3.8]{follmer}.
It means that $h(\nu|\mu)=0$ implies that
the conditional probabilities of $\mu$ coincide with the 
conditional probabilities of $\nu$
for $\nu$-almost all $\omega$. However, as the counterexample 
of \cite{KLR} shows, this result is not suitable
for establishing the variational principle for the 
weakly Gibbsian measures, because the conditional probabilities
can converge to a "wrong" specification. Condition (\ref{newcond}) is instrumental in ensuring that this does not happen.

\section{Almost Gibbs Measures}

In this section we show that the condition (\ref{newcond}) holds for all measures $\nu$ which
are properly supported on the set of continuity points of almost Gibbs specifications.

\begin{defn}\label{spec_alm_gibbs} A specification $\gamma$ is continuous in $\omega$, if for all $\Lambda\in\S$
$$
\sup_{\sigma,\eta}\bigl| \gamma_{\Lambda}(\sigma_\Lambda|
\omega_{\Lambda_n\setminus \Lambda} \eta_{\Lambda_n^c})-
\gamma_{\Lambda}(\sigma_\Lambda| \omega_{\Lambda^c})\bigr| \to
0,\text{ as } n\to\infty.
$$
Denote by $\Omega_\gamma$ the set of all continuity points of
$\gamma$.
\end{defn}

\begin{defn} A measure $\mu$ is called almost Gibbs, if $\mu$ is consistent with a  specification $\gamma$
and
$
\mu(\Omega_\gamma)=1.
$
\end{defn}

\begin{remark} Note that we  define almost Gibbs measures by requiring 
only that the specification is
continuous almost everywhere. We do not require (as it is usually done, 
see e.g. \cite{maesetall})
that the corresponding specification is {\em uniformly non-null}, 
in other words satsfies a finite energy condition: for any
$\Lambda\in\S$, there exist  $a_\Lambda , b_\Lambda \in (0,1)$
such that
$$
a_\Lambda \le \inf_{\xi_{\Lambda},\omega_{\Lambda^c} } \gamma_\Lambda(\xi_{\Lambda}|\omega_{\Lambda^c} ) \le
\sup_{\xi_{\Lambda},\omega_{\Lambda^c} } \gamma_\Lambda(\xi_{\Lambda}|\omega_{\Lambda^c} ) \le b_\Lambda.
$$
\end{remark}

\begin{thm}\label{thm:almostgibbs} If $\mu$ is an almost Gibbs measure for specification $\gamma$, and
$$
\nu(\Omega_\gamma)=1,
$$
then
\begin{equation}\label{newcond1}
\mu(\xi_\Lambda| \omega_{\Lambda_n\setminus\Lambda}) \to \gamma(\xi_\Lambda|\omega_{\Lambda^c})
\end{equation}
for all $\xi_\Lambda$ and $\nu$-almost all $\omega$.
\end{thm}
\begin{proof} Since $\mu\in\G(\gamma)$, $\mu$ satisfies the DLR equations for $\gamma$, and hence
$$
\mu( \xi_\Lambda \omega_{\Lambda_n\setminus\Lambda}) =\int
\gamma_{\Lambda_n} (\xi_\Lambda \omega_{\Lambda_n\setminus\Lambda}|
\eta_{\Lambda_n^c}) \mu(d\eta).
$$
Similarly
\begin{equation}\label{condpr1}
\mu( \xi_\Lambda|
\omega_{\Lambda_n\setminus\Lambda})  = \frac {\mu( \xi_\Lambda
\omega_{\Lambda_n\setminus\Lambda}) }
{\sum_{\tilde\xi_{\Lambda}} \mu( \tilde\xi_\Lambda
\omega_{\Lambda_n\setminus\Lambda}) }=
\frac {\int
\gamma_{\Lambda_n} (\xi_\Lambda \omega_{\Lambda_n\setminus\Lambda}|
\eta_{\Lambda_n^c}) \mu(d\eta)} {\sum_{\tilde\xi_{\Lambda}}\int
\gamma_{\Lambda_n} (\tilde\xi_\Lambda \omega_{\Lambda_n\setminus\Lambda}|
\eta_{\Lambda_n^c}) \mu(d\eta)}.
\end{equation}
Since $\gamma$ is a specification, for all $\xi,\omega,\eta$ one has
\begin{equation}\label{condpr2}
\frac {\gamma_{\Lambda_n} (\xi_\Lambda \omega_{\Lambda_n\setminus\Lambda}|
\eta_{\Lambda_n^c}) } {\sum_{\tilde\xi_{\Lambda}}
\gamma_{\Lambda_n} (\tilde\xi_\Lambda \omega_{\Lambda_n\setminus\Lambda}|
\eta_{\Lambda_n^c}) } = \gamma_\Lambda( \xi_\Lambda | \omega_{\Lambda_n\setminus\Lambda}
\eta_{\Lambda_n^c}) .
\end{equation}
Let
$$
r_{n} (\omega) = \sup_{\xi_\Lambda,\eta_{\Lambda_n\setminus\Lambda}}
|\gamma_\Lambda( \xi_\Lambda | \omega_{\Lambda_n\setminus\Lambda}\eta_{\Lambda_n^c})-\gamma_\Lambda( \xi_\Lambda |
\omega_{\Lambda^c})|.
$$
Therefore, using (\ref{condpr2}), we obtain the following estimate
$$
\gamma_\Lambda(\xi_\Lambda|\omega_{\Lambda^c}) - r_n(\omega)
\le \mu( \xi_\Lambda|
\omega_{\Lambda_n\setminus\Lambda})
\le
\gamma_\Lambda(\xi_\Lambda|\omega_{\Lambda^c}) + r_n(\omega).
$$
Since $r_n(\omega)\to 0$  for  $\omega\in\Omega_\gamma$, we also obtain that
for those $\omega$
$$
 \mu( \xi_\Lambda|
\omega_{\Lambda_n\setminus\Lambda})\to \gamma_\Lambda(\xi_\Lambda|\omega_{\Lambda^c}).
$$
\end{proof}

\begin{remark} In the case $\mu$ is a standard  Gibbs measure and $\gamma$ is the corresponding specification,
 one has $\Omega_\gamma =\Omega$ and hence by repeating  the proof
of Theorem \ref{thm:almostgibbs} one obtains  (\ref{newcond}) for  {\bf all} measures $\nu$.
\end{remark}

\section{Regular Points of Weakly Gibbs Measures}

As we have stressed above the crucial task consists  in determining
the regular (in the sense of (\ref{newcond})) points for $\mu$.
In this section we address this problem in the case of weak Gibbs measures.
A weakly Gibbsian measure $\mu$ is a 
measure for which one can find a potential convergent 
on a set of $\mu$-measure $1$, but not everywhere convergent.

\begin{defn}\label{defn_weakgibbs} Let $\mu$ be a probability measure on 
$(\Omega,\mathcal F)$, and $U=\{U(\Lambda,\cdot)\}$
is an interaction. Then the measure $\mu$ is said to be {\bf weakly Gibbs}
for an interaction $U$ if $\mu$ is consistent with $\gamma^{U}$ 
($\mu\in\mathcal G(\gamma^U)$)
and
$$
\mu(\Omega_{U})=\mu( \{\omega: H_{\Lambda}^U(\omega) \text{ is convergent } 
\forall \Lambda\in\S\})=1.
$$
\end{defn}

Is it natural to expect that the set of points $\Omega_U$ where the potential 
is convergent, coincides with the
set of points regular in the sense of (\ref{newcond}))? 
The definition \ref{defn_weakgibbs} is rather weak.
It is not even clear whether in the case of weak Gibbs measures 
the following convergence holds:
for any finite $\Lambda$, any $\xi_\Lambda$, and $\mu$-almost
 all $\omega$, $\eta$
\begin{equation}\label{spec_conv}
\gamma_\Lambda^U( \xi_\Lambda| \omega_{\Lambda_n\setminus \Lambda}
\eta_{\Lambda_n^c}) \longrightarrow
\gamma_\Lambda^U( \xi_\Lambda| \omega_{\Lambda^c})\ \text{as}\ 
\Lambda_n\uparrow \Z^d.
\end{equation}
 Note that (\ref{spec_conv}) is a natural generalization of
a characteristic property of almost Gibbs measures 
(see definition \ref{spec_alm_gibbs}).

We suspect that (\ref{spec_conv}) does not hold for all weakly Gibbs measures. 
However, the counterexample should be rather pathalogical. Most of the weakly
Gibbs measures known in the literature should be (are) intuitively weak Gibbs
as well.

We introduce a  class of measures which satisfy (\ref{spec_conv}).
\begin{defn}\label{intuitively_wgibbs}
A measure $\mu$ is called {\bf intuitively weakly Gibbs}
for an interaction $U$ if $\mu$ is weakly Gibbs for $U$, and there exists a
a set $\Omega_U^{reg}\subset \Omega_U$ with $\mu(\Omega_U^{reg})=1$ and such that
$$
\gamma_\Lambda^U( \xi_\Lambda| \omega_{\Lambda_n\setminus \Lambda}
\eta_{\Lambda_n^c}) \longrightarrow
\gamma_\Lambda^U( \xi_\Lambda| \omega_{\Lambda^c})
$$
as $\Lambda_n\uparrow \Z^d$, for all $\omega$, $\eta\in
\Omega_{U}^{reg}$.
\end{defn}

This definition of "intuitively" weakly Gibbs measures is
new. However, it  is very natural, and in fact, 
this is how the weakly Gibbs measures have been
viewed before by one of us, c.f. \cite{aernout1}: 
"{\it ... The fact that the constraints which
act as points of discontinuity often involve configurations which
are very untypical for the measure under consideration, suggested
a notion of  {\em almost} Gibbsian or {\em weakly} Gibbsian
measures. These are measures whose conditional probabilities  are
either continuous only on a set of full measure or can be written
in terms of an interaction which is summable only on a set of full
measure. {\em Intuitively}, the difference is that in one case the
``good'' configurations can shield off {\em all} influences from
infinitely far away, and in the other case only {\em almost all}
influences.}

The difference between the Gibbs, almost Gibbs, and intuitively 
weak Gibbs measures is
that
$$
\gamma_\Lambda^U( \xi_\Lambda| \omega_{\Lambda_n\setminus \Lambda}
\eta_{\Lambda_n^c}) \longrightarrow
\gamma_\Lambda^U( \xi_\Lambda| \omega_{\Lambda^c})\ \text{as}\  \Lambda_n\uparrow \Z^d,
$$
holds
\begin{itemize}
\item for {\bf all} $\omega$ and {\bf all} $\eta$ (Gibbs measures);
\item for {\bf $\mu$-almost all} $\omega$ and {\bf all} $\eta$ (almost Gibbs measures);
\item for {\bf $\mu$-almost all} $\omega$ and {\bf $\mu$-almost  all} $\eta$ (intuitively weak  Gibbs measures);
\end{itemize}
A natural question is whether there exists an intuitively 
weak Gibbs which is not almost  Gibbs. An answer is given by the following result.

\begin{thm}\label{classes} Denote by $G$, $AG$, $WG$ and $IWG$ the classes of  Gibbs, 
almost Gibbs,
weakly Gibbs, and intuitively weakly Gibbs states, respectively.
Then
$$
 G \subsetneq AG \subsetneq IWG \subset WG.
$$
\end{thm}

\begin{proof}[Proof of Theorem \ref{classes}] The inclusion $ G \subsetneq AG \subsetneq WG$ was first established in \cite{maesetall}.
The inclusion $IWG\subset WG$ is obvious. In fact, the result of
\cite{maesetall} implies $AG \subset IWG$ as well.

Let us now show that $AG\ne IWG$. In \cite{maesetall}, an example has
been provided of a weakly Gibbs measure, which is not almost
Gibbs. This example is constructed as follows. Let
$\Omega=\{0,1\}^{\Z_+}$ and $\mu$ is absolutely continuous with
respect to the Bernoulli measure $\nu=B(1/2, 1/2)$ with the
density $f$
$$
f(\omega)= \exp(-H^U(\omega)),
$$
where $H^U$ is a Hamiltonian for the interaction $U$, which is
absolutely convergent $\nu$-almost everywhere. The interaction is
defined as follows. Fix $\rho<1$ and define
\begin{equation}\label{U_poten}
U([0,2n],\omega) = \omega_0 \omega_{2n} \rho^{n-N_{2n}(\omega)}
\mathbb I_{\{N_{2n}(\omega)\le n\}},
\end{equation}
where
$$
N_{2n}(\omega) = \max\{ j\ge 1:\,\,
\omega_{2n}\omega_{2n-1}\ldots\omega_{2n-j+1}=1\}
$$
if $\omega_{2n}=1$ and $N_{2n}=0$ if $\omega_{2n}=0$. Moreover,
$U(A,\omega)=0$ if $A\ne[0,2n]$.

It is easy to see that $H^U(\omega)=\sum_{n\ge 0} U([0,2n],\omega)$
 is convergent for $\nu$-a.a. $\omega$.
However, $H^U(\omega)$ is sufficiently divergent, so that the conditional probabilities
$$
\aligned
\mu(\omega_0=1| \omega_1\ldots\omega_n\ldots) &= \frac
{\exp(-H^U(1\omega_1\omega_2\ldots))}{1+\exp(-H^U(1\omega_1\omega_2\ldots))},\\
\mu(\omega_0=0| \omega_1\ldots\omega_n\ldots) &= \frac
{1}{1+\exp(-H^U(1\omega_1\omega_2\ldots))}
\endaligned
$$
are not continuous $\mu$-almost everywhere. Therefore, $\mu$ is not almost Gibbs.

Nevertheless, the exists a set $\Omega'\subset \{0,1\}^{\Z_+}$ such that
$\mu(\Omega')=1$ and for every $\omega,\xi\in \Omega'$
$$
 H^U(1\omega_{0^c})=H^U(1\omega_{[1,2n]}\omega_{[0,2n]^c}), \quad
 H^U(1\omega_{[1,2n]}\xi_{[0,2n]^c}) <\infty,
$$
and
$$
 H^U(1\omega_{[1,2n]}\xi_{[0,2n]^c}) \to H^U(1\omega_{[1,2n]}\omega_{[0,2n]^c}), \quad n\to\infty.
$$
Define
$$
\aligned
B_k &= \{\eta\in\{0,1\}^{\Z_+}: \eta_{2k}\eta_{2k-1}\ldots\eta_{[3k/2]}=1\}, \\
B &= \bigcap_{K\in \mathbb N} \bigcup_{k\ge K} B_k.
\endaligned
$$
Clearly,
$$
\nu(B_k) \le 2^{-k/2}, \quad \text{and}\ \sum_{k} \nu(B_k)<\infty.
$$
Hence, by the Borel-Cantelli lemma $\nu(B)=0$ and since $\mu\ll \nu$, $\mu(B)=0$.

Moreover, for every $\omega \in B^c$, $H^U(\omega)<\infty$.
Let $\Omega' =B^c$ and consider arbitrary $\omega,\xi\in \Omega'$. Then
\begin{align}\label{estimate_difference}
|&H^U(1\omega_{[1,2n]}\xi_{[0,2n]^c})-H^U(1\omega_{[1,\infty)})|\notag\\
&=\bigl|\sum_{p\ge 0} U([0,2p],1\omega_{[1,2n]}\xi_{[0,2n]^c})- U([0,2p],1\omega_{[1,\infty)})\bigr|\notag\\ 
&\le\sum_{p\ge n+1}\bigl| U([0,2p],1\omega_{[1,2n]}\xi_{[0,2n]^c})- U([0,2p],1\omega_{[1,\infty)})\bigr| \\
&\le \sum_{p\ge n+1} U([0,2p],1\omega_{[1,2n]}\xi_{[0,2n]^c})+
\sum_{p\ge n+1} U([0,2p],1\omega_{[1,\infty)})\notag
\end{align}
The sum
$
\sum_{p\ge n+1} U([0,2p],1\omega_{[1,\infty)})
$
converges to zero as $n\to \infty$ since it is a remainder of 
a convergent series for $H^U(1\omega_{[1,\infty)})$.
To complete the proof we have to show that
\begin{equation}\label{sum}
S:=\sum_{p\ge n+1} U([0,2p],1\omega_{[1,2n]}\xi_{[0,2n]^c})
\end{equation}
converges to 0 as well.

For every $\eta\in \Omega'= B^c$ there exists $K=K(\eta)$ such that
$$
\eta_{2k}\eta_{2k-1}\ldots\eta_{[3k/2]}=0
$$
for all $k\ge K$.
Let $K_1=K(\omega), K_2 = K(\xi)$ and $K=\max(K_1,K_2)$.
Suppose $n>K$. Write the sum for $S$ in (\ref{sum}) as $S_1+S_2$, where
$$
S_1=\sum_{p=n+1}^{[4n/3]+1},\quad S_2=\sum_{p=[4n/3]+2}^\infty.
$$
Let us estimate the second sum first.
Since $p>n>\max(K_1, K_2)\ge K(\xi)$, we have that
$$
\xi_{2p}\xi_{2p-1}\ldots \xi_{[3p/2]}=0.
$$
Moreover, since $p\ge [4n/3]+2\ge 4n/3+1$,  one has $[3p/2]>2n$ and therefore
$U([0,2p],1\omega_{[1,2n]}\xi_{[0,2n]^c})$ does not depend on $\omega_{[1,2n]}$.
Hence
$$
S_2 =  \sum_{p=[4n/3]+2}^\infty U([0,2p],1 \xi_{[1,\infty]})  \to 0, \ \text{as } n\to \infty.
$$
Let us now consider the first sum
$$
S_1=\sum_{p=n+1}^{[4n/3]+1} U([0,2p],1\omega_{[1,2n]}\xi_{[0,2n]^c}).
$$
 Terms in $S_1$, in principle, depend on $\omega_{[1,2n]}$. For this
 one has to have that
 $$
   \xi_{2p}\xi_{2p-1} \ldots \xi_{2n+1}=1,
 $$
 and few of the last bits in $\omega_{[1,2n]}$ are also equal to 1. Suppose
 $\omega_{t}=\ldots=\omega_{2n}=1$. Note, however, that since $n>\max(K_1,K_2)\ge K(\omega)$,
 necessarily $t> [3n/2]$. Therefore,
 $$
   U([0,2p],1\omega_{[1,2n]}\xi_{[0,2n]^c}) \le  \rho^{p-(2p-t+1)}
\le\rho^{3n/2 -p-2} \le \rho^{3n/2 -4n/3-3}= \rho^{n/6-3},
 $$
 and since $\rho<1$
 $$
   S_1  \le n\rho^{n/6-3}\to 0.
 $$
\end{proof}

 Another example of an intuitively weakly Gibbs measure, which is not almost Gibbs,
 is the finite absolutely continuous invariant measure of the Manneville--Pomeau
 map \cite{maesetall2}.
 The reason is that for every $\omega$
 $$
 \gamma_{\Lambda}( \omega_0=1| \omega_{[1,n]} {\bf 0}_{[n+1,\infty)} ) =0,
 $$
 where $\bf 0$ is a configuration made entirely from zeros. Thus  the  
 configurations finishing
 with an infinite number of zeros, are the bad configurations, 
 causing the discontinuities in conditional
 probabilities. (One can also show that there are no other such configurations.)
 Since there is at most a countable number of bad configurations,
 this set has a $\mu$-measure equal to $0$.

Yet another example of a measure which should be IWG is the restriction to
a layer of an Ising Gibbs measure.

The reason  for this  is the following.  The example considered above  in Theorem
\ref{classes},   the  absolutely   continuous  invariant   measure   for  the
Manneville-Pomeau map, and the restriction of  an Ising model to a layer, have
a very  similar property  in common. Namely,  for every  "good" configuration
$\omega$, there is a finite number $c=c(\omega)$ such that $ | U(A,\omega)| $
starts to  decay exponentially fast in  $\text{diam}(A)$ as 
soon  as $\text{diam}(A)>c(\omega)$. In
the   example  above,  $c(\omega)   =  K(\omega)$.    In  fact,   the  "good"
configurations  are characterized  by the  property  that $c(\omega)<\infty$.
This random  variable $c(\omega)$ was  called a  {\em correlation length}.
The    main    difficulty    is    in   estimating    the   correlation   length
$c(\xi_\Lambda\omega_{\Lambda_n\setminus\Lambda}\eta_{\Lambda^n})$   for  the
"glued"  configuration
$\xi_\Lambda\omega_{\Lambda_n\setminus\Lambda}\eta_{\Lambda_n^c}$ in terms of
correlations        lengths         $c(\xi_\Lambda\omega_{\Lambda^c})$         and
$c(\xi_\Lambda\eta_{\Lambda^c})$.  Estimates obtained in  
\cite{maes3} should
provide enough information to deal with this problem in case of
 the restriction of the Ising model to a layer.

Let us proceed further with the study of regular 
points of an intuitively weak Gibbs measure $\mu$.
We follow  the proof of Theorem \ref{thm:almostgibbs}. Firstly, one has
\begin{equation}
\frac {\gamma_{\Lambda_n} (\xi_\Lambda \omega_{\Lambda_n\setminus\Lambda}|
\eta_{\Lambda_n^c}) } {\sum_{\tilde\xi_{\Lambda}}
\gamma_{\Lambda_n} (\tilde\xi_\Lambda \omega_{\Lambda_n\setminus\Lambda}|
\eta_{\Lambda_n^c}) } = \gamma_\Lambda( \xi_\Lambda | \omega_{\Lambda_n\setminus\Lambda}
\eta_{\Lambda_n^c}) .
\end{equation}
Let
$$
r_n^\omega(\eta) = \sup_{\xi_\Lambda}\bigl|\gamma_\Lambda( \xi_\Lambda | \omega_{\Lambda_n\setminus\Lambda}
\eta_{\Lambda_n^c})-\gamma_\Lambda( \xi_\Lambda | \omega_{\Lambda^c})\bigl|.
$$
Since $\gamma$ is a specification, $|r_n|\le 2$. 
Moreover, since $\mu$ is intuitively weak Gibbs,
then  for $\omega\in\Omega_U^{reg}$,
$r_n^{\omega}(\eta)\to 0$ for $\mu$-almost all $\eta$. Fix $\eps>0$ and let
$$
 A_{\eps,n}^\omega =\bigl\{\eta:  |r_n^\omega(\eta) | >\eps \bigr\}.
$$
Then
$$
\mu(\xi_\Lambda\omega_{\Lambda_n\setminus\Lambda})= \gamma_\Lambda( \xi_\Lambda | \omega_{\Lambda^c})
\mu(\omega_{\Lambda_n\setminus\Lambda}))
+\int_{\Omega} r_{n}^\omega(\eta) \sum_{\tilde\xi_{\Lambda}}
\gamma_{\Lambda_n} (\tilde\xi_\Lambda \omega_{\Lambda_n\setminus\Lambda}|
\eta_{\Lambda_n^c}) \mu(d\eta),
$$
and we  continue
\begin{align}\label{fractestimate}
\biggl| \frac { \mu(\xi_\Lambda\omega_{\Lambda_n\setminus\Lambda})}
{\mu(\omega_{\Lambda_n\setminus\Lambda})} &-
\gamma_\Lambda( \xi_\Lambda | \omega_{\Lambda^c})
\biggr|\notag\\
&= \biggl| \frac
{(\int_{\Omega\setminus A_{\eps,n}^\omega}+\int_{A_{\eps,n}^\omega}) r_n^\omega(\eta) \sum_{\tilde\xi_{\Lambda}}
\gamma_{\Lambda_n} (\tilde\xi_\Lambda \omega_{\Lambda_n\setminus\Lambda}|
\eta_{\Lambda_n^c}) \mu(d\eta)}
{ \int\sum_{\tilde\xi_{\Lambda}}
\gamma_{\Lambda_n} (\tilde\xi_\Lambda \omega_{\Lambda_n\setminus\Lambda}|
\eta_{\Lambda_n^c}) \mu(d\eta)}\biggr|\\
&\le \eps + 2\frac{
 \int_{A_{\eps,n}^\omega}\sum_{\tilde\xi_{\Lambda}}
\gamma_{\Lambda_n} (\tilde\xi_\Lambda \omega_{\Lambda_n\setminus\Lambda}|
\eta_{\Lambda_n^c})
  \mu(d\eta)}
{ \mu(\omega_{\Lambda_n\setminus\Lambda})},\notag
\end{align}
where we used that $|r_n^\omega|$ is always bounded by $2$.

Let us estimate the remaining integral
$$
\int_{A_{\eps,n}^\omega}\sum_{\tilde\xi_{\Lambda}}
\gamma_{\Lambda_n} (\tilde\xi_\Lambda \omega_{\Lambda_n\setminus\Lambda}|
\eta_{\Lambda_n^c})
  \mu(d\eta) = \int_{\Omega} \mathbb I_{A_{\eps,n}^\omega} (\eta)\bigl( \gamma_{\Lambda_n} \mathbb I_{\omega_{\Lambda_n\setminus\Lambda}}\bigr)(\eta)
\mu(d\eta),
$$
where $ \mathbb I_{\omega_{\Lambda_n\setminus\Lambda}}$ is the indicator of the cylinder  set
$\{\zeta: \zeta_{\Lambda_n\setminus\Lambda}= \omega_{\Lambda_n\setminus\Lambda}\}$.
The set $A_{\eps,n}^\omega$ is $\mathcal F_{\Lambda_n^c}$-measurable, therefore
$$\mathbb I_{A_{\eps,n}} \bigl( \gamma_{\Lambda_n} \mathbb I_{\omega_{\Lambda_n\setminus\Lambda}}\bigr)
= \gamma_{\Lambda_n} \bigl( \mathbb I_{A_{\eps,n}}\mathbb I_{\omega_{\Lambda_n\setminus\Lambda}}\bigr)
$$
and since $\mu$ satisfies the DLR equations with $\gamma$, we obtain that
\begin{align*}
\int_{A_{\eps,n}}\sum_{\tilde\xi_{\Lambda}}
\gamma_{\Lambda_n} (\tilde\xi_\Lambda \omega_{\Lambda_n\setminus\Lambda}|
\eta_{\Lambda_n^c})
  \mu(d\eta)& =\int\mathbb I_{A_{\eps,n}^\omega}(\eta)\mathbb I_{\omega_{\Lambda_n\setminus\Lambda}}(\eta) \mu(d\eta)\\
&=
\mu( A_{\eps,n}^\omega \cap \omega_{\Lambda_n\setminus\Lambda}).
\end{align*}
Therefore, we obtain the following estimate
$$
\biggl| \frac { \mu(\xi_\Lambda\omega_{\Lambda_n\setminus\Lambda})}
{\mu(\omega_{\Lambda_n\setminus\Lambda})} -
\gamma_\Lambda( \xi_\Lambda | \omega_{\Lambda^c})
\biggr| \le \eps + 2\frac { \mu(\omega_{\Lambda_n\setminus\Lambda}\cap A_{\eps,n}^\omega)}
{ \mu(\omega_{\Lambda_n\setminus\Lambda})}.
$$
Now, if we can show that for all $\omega\in\Omega_U^{reg}$
$$
\frac { \mu(\omega_{\Lambda_n\setminus\Lambda}\cap A_{\eps,n}^\omega)}
{ \mu(\omega_{\Lambda_n\setminus\Lambda})}\to 0,\ \text{as}\ \Lambda_n\uparrow\mathbb Z^d,
$$
we will be able to conclude that all points $\Omega_U^{reg}$ are regular in the sense of
(\ref{newcond}).

Let us now turn to the example of an intuitively weak Gibbs measure considered in Theorem \ref{classes}.
As usual for the Gibbs formalism, we check the required property only for $\Lambda=\{0\}$. We also let $\Lambda_n=[0,n]$,
hence $\Lambda_n\setminus\Lambda=[1,n]$.

Let $\omega\in \Omega_U^{reg}$. Hence the potential is convergent in
${\bf 1_0}\omega_{[1,\infty)}$, and therefore
$\exp(-H({\bf 1_0}\omega_{[1,\infty)}))>0$. Choose arbitrary $\eps>0$ such that
$$
\eps < \frac 14 \exp(-H({\bf 1_0}\omega_{[1,\infty)})).
$$
The measure $\mu$ is absolutely continuous with respect to the Bernoulli measure $\nu=B(1/2,1/2)$.
First of all, let us show that
\begin{equation}\label{nu_estimate}
\frac { \nu(\omega_{[1,n]}\cap A_{\eps,n}^\omega)}
{ \nu(\omega_{[1,n]})}\to 0,\ \text{as}\ n\to\infty,
\end{equation}
implies
\begin{equation}\label{mu_estimate}
\frac { \mu(\omega_{[1,n]}\cap A_{\eps,n}^\omega)}
{ \mu(\omega_{[1,n]})} \to 0,\ \text{as}\ n\to\infty,
\end{equation}
Since $H(\zeta)$ is non-negative (possibly infinite) for any $\zeta$, one has
$$\aligned
\mu( \omega_{\Lambda_n\setminus\Lambda}\cap A_{\eps,n}^\omega) &=\int\limits_{\omega_{\Lambda_n\setminus\Lambda}\cap A_{\eps,n}^\omega}
\exp( -H(\zeta)) \nu( d\zeta) \\
&\le \int\limits_{\omega_{\Lambda_n\setminus\Lambda}\cap A_{\eps,n}^\omega}
\nu( d\zeta) =\nu( \omega_{\Lambda_n\setminus\Lambda}\cap A_{\eps,n}^\omega).
\endaligned
$$
Consider the  set  $W=\omega_{[1,n]}\cap (A_{\eps,n}^\omega)^c$.
For every $\eta\in W$ we have
$$
\sup_{\xi_0}\Bigl|\gamma_0(\xi_0|\omega_{[1,n]}\eta_{[n+1,\infty)}) -
\gamma_0(\xi_0|\omega_{[1,\infty)})
\Bigr|\le \eps.
$$
In particular
$$
\Biggl|
\frac
{\exp\bigl(-H({\bf 1}_0\omega_{[1,n]}\eta_{[n+1,\infty)})\bigr)}
{
\exp\bigl(-H({\bf 1}_0\omega_{[1,n]}\eta_{[n+1,\infty)})\bigr)
+
1
}-
\frac
{\exp\bigl(-H({\bf 1}_0\omega_{[1,\infty)})\bigr)}
{
\exp\bigl(-H({\bf 1}_0\omega_{[1,\infty)})\bigr)
+
1 
}
\Biggr|\le \eps,
$$
where we have used the fact that $H({\bf 0}_0\zeta_{[1,\infty)})=0$ for all
$\zeta$. Note also that since $\omega$, $\eta\in\Omega_U^{reg}$, both
$H({\bf 1}_0\omega_{[1,\infty)})$,
$H({\bf 1}_0\omega_{[1,n]}\eta_{[n+1,\infty)})$ are non-negative and finite.
Therefore
$$
\Bigl|
{\exp\bigl(-H({\bf 1}_0\omega_{[1,n]}\eta_{[n+1,\infty)})\bigr)}
-
{\exp\bigl(-H({\bf 1}_0\omega_{[1,\infty)})\bigr)}
\Bigr|\le 4\eps.
$$
Hence,
$$\aligned
\mu( \omega_{[1,n]})& \ge \mu( \omega_{[1,n]}\cap (A_{\eps,n}^{\omega})^c )\\
&= \mu({\bf 1}_0 \cap\omega_{[1,n]}\cap (A_{\eps,n}^{\omega})^c )+
 \mu({\bf 0}_0 \cap\omega_{[1,n]}\cap (A_{\eps,n}^{\omega})^c )\\
&= \int\limits_{{\bf 1}_0\cap \omega_{[1,n]}\cap(A_{\eps,n}^{\omega})^c  } \exp(-H(\zeta))\,
\nu(d\zeta)\\
&\phantom{\quad\quad\quad}+
 \int\limits_{{\bf 0}_0\cap \omega_{[1,n]}\cap(A_{\eps,n}^{\omega})^c  } \exp(-H(\zeta))\,
\nu(d\zeta)\\
&\ge \Bigl( \exp( -H({\bf 1}_0 \omega_{[1,\infty)})-4\eps\Bigr)
\nu( {{\bf 1}_0\cap \omega_{[1,n]}\cap(A_{\eps,n}^{\omega})^c })\\
&\phantom{\quad\quad\quad}+
 \nu( {\bf 0}_0\cap \omega_{[1,n]}\cap(A_{\eps,n}^{\omega})^c )\\
&\ge C
\nu( \omega_{[1,n]}\cap(A_{\eps,n}^{\omega})^c ),
\endaligned
$$
where $C= \exp( -H({\bf 1}_0 \omega_{[1,\infty)})-4\eps>0$ (note that $C<1$).
Therefore,
$$\aligned
\frac { \mu(\omega_{[1,n]}\cap A_{\eps,n}^\omega)}
{ \mu(\omega_{[1,n]})} &\le C^{-1} \frac { \nu(\omega_{[1,n]}\cap A_{\eps,n}^\omega)}
{ \nu(\omega_{[1,n]}\cap(A_{\eps,n}^{\omega})^c)}\\
&=C^{-1} \frac { \nu(\omega_{[1,n]}\cap A_{\eps,n}^\omega)}
{ \nu(\omega_{[1,n]})}
\frac 1{1-\frac { \nu(\omega_{[1,n]}\cap A_{\eps,n}^\omega)}
{ \nu(\omega_{[1,n]})}},
\endaligned
$$
and hence (\ref{nu_estimate}) indeed implies (\ref{mu_estimate}).

Let us now proceed with the proof of (\ref{nu_estimate}). Since $\nu$ is a symmetric
Bernoulli measure, $\nu(\omega_{[1,n]}) =2^{-n}$.

If $x,y\ge 0$ then
$$
\Biggl| \frac {e^{-x}}{1+e^{-x}}- \frac {e^{-y}}{1+e^{-y}}\Biggr|=
\Biggl| \frac {1}{1+e^{-x}}- \frac {1}{1+e^{-y}}\Biggr|\le |x-y|.
$$
Therefore, if $\eta \in A_{\eps,n}^\omega$, i.e.,
$$
\sup_{\xi_0}\Bigl|\gamma_0(\xi_0|\omega_{[1,n]}\eta_{[n+1,\infty)}) -
\gamma_0(\xi_0|\omega_{[1,\infty)})
\Bigr|> \eps,
$$
then
\begin{equation}\label{B_set}
\Bigl| H({\bf 1}_0\omega_{[1,n]}\eta_{[n+1,\infty)})-
H({\bf 1}_0\omega_{[1,\infty)})\Bigr|>\eps.
\end{equation}
Hence, if we define $B_{\eps,n}^{\omega}$ as a set of points $\eta$ such that
(\ref{B_set}) holds, we get that $A_{\eps,n}^{\omega}\subseteq B_{\eps,n}^{\omega}$.

To estimate the measure of $B_{\eps,n}^{\omega}$ we have to use
the estimates from the proof of Theorem \ref{classes}.
Without loss of generality we may assume that $n$ is even,
$n=2n'$. Let us recall the estimate (\ref{estimate_difference})
\begin{align}\label{estimate_difference1}
|&H({\bf 1}_0\omega_{[1,2n']}\eta_{[2n'+1,\infty)})-H({\bf 1}_0\omega_{[1,\infty)})|\notag\\
&\le \sum_{p\ge n+1} U([0,2p],{\bf 1}_0\omega_{[1,2n']}\eta_{[2n'+1,\infty)})+ 
\sum_{p\ge n'+1} U([0,2p],{\bf
1}_0\omega_{[1,\infty)})\notag
\end{align}

The second sum on the right hand side does not depend on $\eta$, 
and converges to $0$ as $n'\to\infty$.
Therefore, by choosing $n'$ large enough we must have that if 
$\eta \in B_{\eps,n}^\omega$ then
$$
 \sum_{p\ge n+1} U([0,2p],{\bf 1}_0\omega_{[1,2n']}\eta_{[2n'+1,\infty)})> \frac {\eps}{2}.
$$

Let us define a sequence $\delta_p =\rho^{0.1p}$, $p\ge 1$.
Since $\rho\in (0,1)$,  for sufficiently large $n'$ one has
$$
\sum_{p\ge n'+1} \delta_p < \frac {\eps}{2}.
$$
Consider the following events, 
$$ 
  C_{p}^{\omega} = \Bigl\{\eta:  U([0,2p], 
{\bf 1}_0\omega_{[1,2n']}\eta_{[2n'+1,\infty)}) > \delta_p \Bigr\}.
$$
Obviously, 
$$B_{n,\eps}^{\omega} \subset \bigcup_{p\ge n'+1} 
C_{p}^{\omega}.
$$
In general, for arbitrary $\zeta$, $U([0,2p],\zeta)> \delta_p$ if 
(see (\ref{U_poten})) $\zeta_0=\zeta_{2p}=1$,
$N_{2p}(\zeta)\le p$ and 
and 
$
\rho^{ p-N_{2p}(\zeta)} > \rho^{0.1p}$. Therefore,
$$
  0.9p \le  N_{2p}(\zeta) \le p,
$$
and hence
$$
\nu( \zeta: U([0,2p],\zeta)>\delta_p) \le 
\sum_{k=[0.9p]}^p 2^{-k}\le 2^{-0.9p+2}=: z_p.
$$
Let us continue with estimating the probability of 
$C_{p}^{\omega}$. 
If $p> 2n'$, then $C_{p}^{\omega}$ 
does not depend on $\omega$, and hence using the previous estimate
$$
\nu(\omega_{[1,2n']}\cap C_{p}^{\omega}) \le 
2^{-2n'} z_p.
$$

For the small values of  $p$,  $p\in [n'+1,2n']$, we have to
proceed 
differently. For such $p$'s the configuration $\eta$ can ``profit''
from the last bits (equal to 1) in $\omega$. Since $\omega$ is 
a regular configuration (see Theorem \ref{classes}), 
for sufficiently large $n'$, 
$\omega \in G_{n'}$, where 
$$
  G_{n'} 
=\{\zeta: \zeta_{2k}\ldots \zeta_{[3k/2]}=0\quad\forall k\ge n'\}.
$$
In particular, it means that at most $n'/2+1$ of the last bits in 
$\omega_{[1,2n']}$  are equal to $1$, and in the worst case, 
$\omega_{2n'}\ldots \omega_{[3n'/2]+1} =1$. From now
on we assume that $\omega_{2n'}\ldots \omega_{[3n'/2]+1} =1$.

We split the set of ``bad'' $\eta$'s as follows:
$$\aligned
C_p^{\omega}=\{\eta:\, & U([0,2p],{\bf 1}_0\omega_{[1,2n']}\eta_{[2n'+1,\infty)})
>\delta_p \}\\
& =\{\eta: U([0,2p],{\bf 1}_0\omega_{[1,2n']}\eta_{[2n'+1,\infty)})
>\delta_p\,\&\, \eta_{2n'+1}\ldots\eta_{2p}=0 \}\cup \\
&\phantom{quad}\{\eta: U([0,2p],{\bf 1}_0\omega_{[1,2n']}\eta_{[2n'+1,\infty)})
>\delta_p\,\&\, \eta_{2n'+1}\ldots\eta_{2p} =1 \}\\
&= C_p^{\omega,0}\cup C_p^{\omega,1}.
\endaligned
$$
Again, the set $ C_p^{\omega,0}$  
does not depend on $\omega$. In fact, $ C_p^{\omega,0}$  
is not empty only for $p$'s close to $2n'$: on one hand,
$p-N_{2p}<0.1p$ and on the other, $N_{2p}\le 2p-2n'$. 
Together with the fact that $p\in (n',2n']$, this is possible
only 
for $\displaystyle p'\in \Bigl( \frac 2{1.1}n', 2n'\Bigr]$.
For any $p$ in this interval, one would need more than $0.9p$ ones,
hence making a $\nu$-measure of $C_p^{\omega,0}$ sufficiently small: 
$$
\nu(C_p^{\omega,0} 
) \le 2^{-0.9p}.
$$ 

%
Finally, the elements of $C_p^{\omega,1}$ 
are precisely the configurations
which can profit from the fact that the last few bits in 
$\omega_{[1,2n']}$ are equal to $1$.
For such $\eta$'s, in a ``glued'' 
configuration $\zeta = {\bf 1}_0\omega_{[1,2n']}
\eta_{[2n'+1,\infty)}$  a continuous interval of $1$'s
is located   starting from position $[3n'/2]+1$ and
finishing at position $2p$. 
In order to have a positive contribution from
$U([0,2p],\zeta)$ a
long run of $1$'s should not be too long. Namely,
$$
N_{2p}(\zeta) = 2p - \Bigl[\frac {3n'}{2}\Bigr]\le p,
$$
implying that $p\le [3n'/2]$, and hence, $C_{p}^{\omega,1}$ is 
empty for $p>[3n'/2]$. 
For, $p\in [n'+1, [3n'/2] ]$ one has
$$
U([0,2p],\zeta) = \rho^{p-N_{2p}(\zeta)} = \rho^{ [3n'/2]-p}.
$$
Hence if $U([0,2p],\zeta)>\rho^{0.1p}$, then $[3n'/2]-p<0.1p$ and
hence $p> [3n'/2]/1.1$. Once again that means that $C_{p}^{\omega,1}$
is empty for $p\in [n'+1, [3n'/2]/1.1-1]$. 

Therefore, for $p\in [n'+1,2n']$ we conclude that
$$
C_p^{\omega,1} \subset \{\eta: \eta_{2n'+1} =\ldots =\eta_{2p} =1\}
\quad\text{if}\quad  \frac 1{1.1} \Bigl[\frac {3n'}{2}\Bigr]<p\le 
\Bigl[\frac {3n'}{2}\Bigr],
$$
and $  C_p^{\omega,1} =\varnothing$, otherwise. In any case,
$$
\nu(C_p^{\omega,1}) \le 2^{-2p+2n'}.
$$

We obtained that
$$\aligned
\nu( \omega_{[1,2n']}\cap A_{\eps,n}^\omega) &\le 
\nu( \omega_{[1,2n']}\cap B_{\eps,n}^\omega) \le
\nu\Bigl( \omega_{[1,2n']}\cap \cup_{p\ge n'+1}C_p^\omega\Bigr)\\
&\le \sum_{p\ge n'+1} \nu( \omega_{[1,2n']} \cap C_p^\omega)= 
S_1+S_2+S_3+S_4,
\endaligned
$$
where $S_1$, $S_2$, $S_3$, $S_4$ are sums over integer $p$'s
in intervals 
$ I_1=[n'+1, [3n'/2]/1.1)$, $I_2=[ [3n'/2]/1.1, [3n'/2]]$,
$ I_3=[ [3n'/2]+1, 2n']$, and $I_4=[2n'+1,\infty)$, respectively.
We have the following estimates
$$\aligned
S_1 & =\sum_{p\in I_1} \nu(\omega_{[1,2n']}\cap C_p^{\omega} )
     =\sum_{p\in I_1} \nu(\omega_{[1,2n']}\cap C_p^{\omega,0} )\\
    & =\sum_{p\in I_1} \nu(\omega_{[1,2n']})\nu(C_p^{\omega,0} ) 
\le 2^{-2n'} \sum_{p\in I_1} 2^{-0.9p} \\&\le  2^{-2n'}\frac {2^{-0.9n'}} {1-2^{-0.9}}
\le 3\cdot 2^{-2.9 n'};\\
S_2 &=\sum_{p\in I_2} \nu(\omega_{[1,2n']}\cap C_p^{\omega,0} )+
\sum_{p\in I_2} \nu(\omega_{[1,2n']}\cap C_p^{\omega,1} )\\
 & \le 2^{-2n'}\sum_{p\in I_2} 2^{-0.9p} +2^{-2n'} \sum_{p\in I_2} 2^{-2p+2n'}\\
&\le 2^{-2n'}\cdot 12\cdot 
2^{- \frac {0.9}{1.1} \cdot \frac {3n'}{2}} 
+12\cdot 2^{-\frac 2{1.1}\cdot \frac {3n'}{2}}
 \\
&\le 12\cdot 2^{-3n'}+12\cdot 2^{-2.7n'}\le 12\cdot 2^{-2.7n'};\\
S_3 &= \sum_{p\in I_3} \nu(\omega_{[1,2n']}\cap C_p^{\omega} )
     =\sum_{p\in I_3} \nu(\omega_{[1,2n']}\cap C_p^{\omega,0} )\\
    & =2^{-2n'} \sum_{p\in I_3} 2^{-0.9p} \le  
2^{-2n'}\cdot 12\cdot 2^{-\frac {0.9\cdot 3n'}{2} }\le 
12\cdot 2^{-3n'};\\
S_4 &=  \sum_{p\in I_4} \nu(\omega_{[1,2n']}\cap C_p^{\omega} )\le 12\cdot 
2^{-3.8n'}.
\endaligned
$$
Finally, we conclude that 
$$
\frac {\nu( \omega_{[1,2n']}\cap A_{\eps,n}^\omega)} {\nu(\omega_{[1,2n']})} \le
\frac {S_1+S_2+S_3+S_4}{2^{-2n'}} \to 0\text{ as }n'\to\infty. 
$$

To summarize our result, we formulate the following theorem.

\begin{thm} Let $\mu$ be the (intuitively) weak Gibbs measure, 
but not almost Gibbs, discussed
above in  Theorem \ref{classes}, and which has been 
introduced in \cite{maesetall}.
Then there exists a set $\Omega'$ such 
that $\mu(\Omega')=1$ and the following holds:
\begin{itemize}
\item the potential $U$ is absolutely convergent on $\Omega'$;
\item for all $\omega$, $\eta\in\Omega'$,
any finite $\Lambda$ and all $\xi_\Lambda\in \Omega_\Lambda$ one has
$$
   H_\Lambda(\xi_\Lambda\omega_{\Lambda_n\setminus \Lambda}\eta_{\Lambda_n^c})\to 
   H_\Lambda(\xi_\Lambda\omega_{\Lambda^c}),
$$
$$
 \gamma_\Lambda(\xi_\Lambda|\omega_{\Lambda_n\setminus \Lambda}\eta_{\Lambda_n^c})\to 
   \gamma_\Lambda(\xi_\Lambda|\omega_{\Lambda^c}),
$$
as $\Lambda_n\to\mathbb Z_{+}$.
\item every $\omega\in\Omega'$ is regular in (Goldstein's) sense: for every $\xi_\Lambda$
$$
\mu(\xi_\Lambda| \omega_{\Lambda_n\setminus\Lambda}) \to   \gamma^U_\Lambda(\xi_\Lambda|\omega_{\Lambda^c}),
$$
as  $\Lambda_n\to\mathbb Z_{+}$.
\end{itemize}  
\end{thm}
\section{ Bit-shift channel}
A somewhat  different kind  of non-Gibbsian example
comes from an industrial application: data storage on  magnetic tape or 
optical disks (CD, DVD, etc).   Before formulating the model precisely, 
let us explain the mechanism which leads to a non-Gibbsian measure.

The medium for  magnetic or optical data  storage can be in one 
of the two states:
``high'' and ``low'', or ``bright'' and ``dark''.  The information is encoded
not in the  state of  the medium  itself, but  in transitions  between these
states, and more precisely, in ``units of time'' between two successive  transitions.
In the following table the first line indicates the state of the medium (H(igh) or L(ow))
and the second line  indicates the corresponding occurrence (1) or absence (0) of transitions:
\begin{center}
\begin{tabular}{lc@{}c@{}c@{}c@{}c@{}c@{}c@{}c@{}c@{}c@{}c@{}c@{}c@{}c@{}c@{}c@{}c@{}c@{}c@{}c@{}c@{}c@{}c@{}c@{}c@{}cr}
$\ldots$&L& &H& &H& &H& &H& &L& &L& &L& &H& &H& &H& &H& &L& &$\ldots$ \\
$\ldots$& &1& &0& &0& &0& &0& &1& &0& &1& &0& &0& &0& &1& & &$\ldots$ \\
\end{tabular}
\end{center}
An equivalent way to represent the second line is to record the number of 
zeros between consecutive ones. In the case above, one obtains a sequence
$(\ldots, 3,2,3,\ldots)$. For technical reasons, in data storage 
one often uses coding schemes such  that the transitions are never too close,
but also not too far away from each other. This is achieved by using
the so-called {\em run-length constrained codes}.

When the magnetic medium or optical disk are read, due to various  effects
like noise, intersymbol interference or  clock  jittering,  the
transitions can  be erroneously  identified, thus producing  a time-shift  in the
detected  positions. 

Suppose in the example above the following error has occurred:
the second transition has been detected one time unit too late. The resulting 
sequence then is $(\ldots, 1, 0, 0, 0, 0,1, 0, 1, 0 ,0, 0, 1,\ldots)$.
And the corresponding representation in terms of runs of zeros will be
 $(\ldots, 4,1,3,\ldots)$ instead of $(\ldots,3,2,3,\ldots)$.

The following description of a bit shift channel 
is due to Shamai and Zehavi, \cite{shamai}.

Let $\mathcal A=\{d,\ldots,k\}$, where $d,k\in \mathbb N$, $d<k$ and $d\ge 2$.
Define $X=\mathcal A^\mathbb Z=\{ x=(x_i): x_i\in \mathcal A\}$, 
$\Omega=\{-1,0,1\}^\mathbb Z=\{ \omega=(\omega_i): \omega_i\in \{-1,0,1\} \}$.
Consider the following transformation $\phi$ defined on $X\times \Omega$ as follows:
$ y =\phi( x, \omega)$
with
$$
y_i = x_i+\omega_i-\omega_{i-1}\quad \text{for all}\quad i\in \mathbb Z.
$$
Note that $ y$ is a sequence  such that $y_i\in \{0,\ldots, k+2\}$ for all $i$,
but not every sequence in $\{0,\ldots,k+2\}^\mathbb Z$ can be obtained
as an image of some $ x\in X$, $ \omega\in\Omega$. For example, 
all image sequences $y=\phi(x,d)$ cannot contain $00$. Indeed, 
suppose $y_i=0$ for some $i$. 
This is possible if and only if  $x_i=2$, $\omega_i=-1$, and $\omega_{i-1}=1$. 
But then $y_{i+1}=x_{i+1}+\omega_{i+1}-\omega_{i}\ge 2-1+1=2$.

Since $\phi$ is a continuous (in the product topology) transformation
the set $Y=\phi(X\times\Omega)$ is a so-called sofic shift, see \cite{symb}.

Suppose $\mu$ and $\pi$ are product Bernoulli measure on $X$ and $\Omega$ with
$$
  \mu( j )=p_j,\ j=d,\ldots,k, \  \pi(-1)=\pi(1)=\epsilon,\ 
\pi(0)=1-2\epsilon.
$$
The measure $\mu$ describes the source of information and 
$\pi$  describes the jitter (noise).

Let $\nu=
(\mu\times\pi)\circ\phi^{-1}$ be a corresponding factor measure on $Y$ defined by
$$
\nu(C) := (\mu\times\pi) \bigl(\phi^{-1}C\bigr) 
\text{ for any Borel measurable }  A\subset Y.
$$

Despite the fact that some configurations are forbidden in $Y$,
in other words, we have some ``hard-core'' constraints, there is a rich  
theory of Gibbs measures 
for sofic subshifts.
One of the equivalent ways  to define Gibbs measures is as follows.
We say that an invariant measure $\rho$ on $Y$  is  Gibbs for a H\"older 
continuous function $\phi:Y\to \mathbb R$ and constants $P$ and $C>1$ such that
 for any $y\in Y$ one has
\begin{equation}\label{pot_psi1}
C^{-1} \le \frac {\rho([y_0,y_1,\ldots,y_n])}
{\exp\bigl( \sum_{k=0}^n \phi( \sigma^k y) - (n+1)P\bigr)}\le C,
\end{equation}
where $\sigma:Y\to Y$ is the left shift.The function $\phi$ is often called a 
{\em potential}, and has a
role  analogous to that of $f_U(\cdot) 
=\sum_{0\in A} U(A,\cdot)/|A|$ for standard lattice systems. The constant  
$P$ in (\ref{pot_psi1}) is in fact the pressure of $\phi$.

Now, (\ref{pot_psi1}), often called the {\em Bowen-Gibbs} property, implies that
for all $y\in Y$
\begin{equation}\label{pot_psi2}
\phi(y)-C_1 \le \log \rho(y_0|y_1,\ldots,y_n) \le \phi(y)+C_1,
\end{equation}
for some positive constant $C_1$. Since $Y$ is compact, and $\phi$ is continuous,
we conclude that for every $y$ and all $n\in\mathbb N$ the logarithm 
of the conditional probability  $\rho(y_0|y_1,\ldots,y_n)$ is bounded
from below and above.

It turns out that $\nu=(\mu\times\pi)\circ\phi^{-1}$ is not Gibbs. 
As usual in the study  of non-Gibbsianity we have to indicate 
a bad configuration.
In our case, configuration $02^\infty$  is a bad configuration for $\nu$.
Consider  cylinder $[y_0,\ldots,y_n]$ where
$$
 y_0=0,\ y_1=\ldots =y_n=2
$$
Then effectively there is a unique preimage of this cylinder.
Indeed $y_0=0$, and as we have seen above, this is possible only 
for 
$$
x_0 =2,\ \omega_0 = -1,\ \omega_{-1} = 1.
$$
For the next position $i=1$ we have
$$
2= y_1 = x_1 +\omega_1-\omega_{0} = x_1+\omega_1+1. 
$$
Again, since $\omega_1+1\ge 0$ and $x_1\ge 2$, this is possible if and only
if $\omega_1=-1$ and $x_1=2$. But then $x_2=2$ and $\omega_2=-1$, and so on.
Therefore
$$
\phi^{-1}( [0,\underbrace{2,2,\ldots, 2}_{n \text{ times}}]) \subset 
[\underbrace{2,2,\ldots, 2}_{n+1 \text{ times}} ]\times 
[\underbrace{-1, -1,\ldots, -1}_{n+1 \text{ times}}],
$$
and hence
$$
\nu([0,2,2,\ldots,2]) \le \mu([2,2,\ldots,2])\pi([-1,-1,\ldots,-1])= 
(p_2\epsilon)^{n+1}.
$$
On the other hand, cylinder $[2,2,\ldots,2]$ has many preimages.
For example, with appropriate choice of $\omega$'s cylinders of
the form 
$$
[x_1,\ldots,x_n]= [2,\ldots,2,3,2,\ldots,2]\subset X
$$ 
will  project into $[2,\ldots,2]$. Indeed, if $j$ is  the position of $3$ in 
$[x_1,\ldots,x_n]$, then the choice  $\omega_{0}=\omega_1=\ldots=\omega_{j-1} =0$, and
$\omega_j=\omega_{j+1}=\ldots=\omega_{n}=-1$ will suffice.
Therefore
$$
\nu([2,2,\ldots, 2]) 
\ge \sum_{j=1}^n p_2^{n-1}p_3 (1-2\epsilon)^{j} \epsilon^{n-j+1}, 
$$
and for $\epsilon<1/3$, one has
$$
\nu([2,2,\ldots, 2]) \ge n p_2^{n-1} p_3 \epsilon^{n+1},
$$
and therefore
$$
\nu(0|2,2,\ldots,2)=\frac { \nu([0,2,2,\ldots,2])}{\nu([2,2,\ldots,2])}\le\frac C{n},
$$
and thus the logarithm of $\nu(0|2,2,\ldots,2)$
is not uniformly bounded from below, 
and hence there is no H\"older continuous $\phi$ such that
 (\ref{pot_psi2}) is valid for $\nu$,
 and hence, $\nu$ is not Gibbs.

A slightly more accurate analysis shows that $\nu$ is not Gibbs for 
$\epsilon>1/3$ as well.

An  interesting open problem is the computation of
the capacity of the bit-shift channel with a fixed jitter measure $\pi$.
This problem reduces to the computation of the
entropy  of the transformed   measure $\nu$ for an arbitrary input measure $\mu$. 
In \cite{stan} an efficient algorithm 
was proposed for  Bernoulli measures $\mu$.
This algorithm produces  accurate (to arbitrary precision)
numerical lower and upper bounds
on the entropy of $\nu$.


\section{ Discussion}
In this paper we addressed the problem of finding  sufficient conditions
under which  $h(\nu|\mu)=0$ implies that $\nu$ is consistent with a given 
specification $\gamma$ for $\mu$. In particular, the question is  interesting 
in the case of an almost or a weakly Gibbs measure $\mu$.
 Intuition developed in \cite{ES,maesetall2,KLR} shows that $\nu$ must be 
concentrated on a set of ``good'' configurations for measure $\mu$. 
In the case $\mu$ is almost Gibbs, $\nu$ must be concentrated on
the continuity points $\Omega_\gamma$, \cite{KLR}.
A natural generalization to the case of a weakly Gibbs measure
$\mu$ for potential $U$ would be to assume that 
$\nu$ is concentrated on the  convergence points of the Hamiltonian $H^U$.
However, this is not true as the counterexample of \cite{KLR} shows. 

We weakened and generalized the conditions under which we can prove the first 
part of the  Variational  Principle. Moreover, we introduced the class of 
Intuitively  Weak Gibbs measures, which is strictly larger than the almost 
Gibbs class,  but contained in  the Weak Gibbs class.

The example considered in this paper shows (and we conjecture
the same type of behaviour for other interesting examples of weakly 
Gibbs measures) that some weak Gibbs measures are more regular 
than was thought before.


\end{document}